\documentclass[aps,preprint,showpacs,preprintnumbers,amsmath,amssymb]{revtex4}

\begin{document}

\title{Amplitude of Coulomb scattering for charged scalar field in de Sitter spacetime}
\author{Crucean Cosmin}
 \email{crucean@quasar.physics.uvt.ro}
 \affiliation{West University of Timi\c soara,  V. Parvan \\
 Avenue 4 RO-300223 Timi\c soara,  Romania}

\date{\today}

\begin{abstract}
The scattering of a charged scalar field on Coulomb potential is studied using solutions of the Klein-Gordon equation which have a definite momentum. One obtains that in contrast with what happens on Minkowski case the modulus of momentum is not conserved in the scattering process on de Sitter space. 
\end{abstract}


\maketitle

\section{\label{sec:1}Introduction}

The quantum scalar field is studied in problems concerning quantum
effects in the presence of gravitation and cosmology. Of a
special interest in cosmology is the de Sitter expanding universe carrying
scalar fields variously coupled with gravitation. In the quantum theory a
special role is played by the free fields (minimally coupled to gravitation)
because with their help we can calculate scattering amplitudes
using perturbation methods. 

The theory of quantum scalar field on de Sitter space was intensively studied, but the explicit solutions were written in a few papers. In our opinion these solutions have a crucial importance, in developing the theory of interaction on de Sitter space-time. We focus our attention on this subject because the last astronomical observations shows that the expansion of the Universe is accelerating, thus increasing the interest on de Sitter model. The interacting theory in scalar electrodynamics on de Sitter spacetime in the present literature is concerned with important effects such as: vacuum polarization, self mass and stress-energy tensor, but neglect the scattering processes. This theory of scattering could be of interest, since the study of interactions between fermions, bosons and scalar fields on a curved background will help us to understand how the well established picture from flat space will be modified by a gravitational field. Until now detailed studies related to the Dirac field \cite{B2},\cite{B16},\cite{B14},\cite{B15} and scalar field \cite{B18},\cite{B8} on de Sitter space-time was done obtaining important results such as normalized analytical solutions and quantization. The less studied is the electromagnetic field, but here the situation is simple if one recalls the conformal invariance between de Sitter and Minkowski metrics. Thus at this moment seems that we have all elements for constructing a theory of interaction between quantum fields on de Sitter background. In this paper our attention will be focused on the interaction between scalar field and electromagnetic field.

We start in the second section with a brief review of the main results concerning the quantum scalar field theory on de Sitter space. Section \ref{sec:3} is dedicated to the study of Coulomb scattering of a charged scalar particle on de Sitter space. In section \ref{sec:4} we study the problems that arise when we pass from scattering amplitude to cross section. Our conclusions are summarized in section \ref{sec:5}, and in Appendix we give the main steps leading to our scattering amplitude. These results are presented in natural units $\hbar=c=1$.

\section{\label{sec:2}Plane wave solutions for scalar field}

We start with the exact solutions of the free Klein-Gordon equation on de Sitter space-time, which was obtained in \cite{B8},\cite{B12}. Let us write the de Sitter line element \cite{B1}:
\begin{equation}
ds^{2}=dt^{2}-e^{2\omega t}d\vec{x}^{\,2},\label{eq:m}
\end{equation}
where $\omega$ is the expansion factor.
In the chart $\{t,\vec{x}\}$ with Cartesian coordinates the Klein-Gordon
equation can be analytically solved
in terms of Bessel functions \cite{B18},\cite{B8},\cite{B9}. From
different versions of solutions which are currently used, we will use here the
normalized solutions of positive frequencies from \cite{B12}:
\begin{equation}
f_{\vec{
p}}\,(x)=\frac{1}{2}\sqrt{\frac{\pi}{\omega}}\,\frac{e^{-3\omega
t/2}}{(2\pi)^{3/2}}\,e^{-\pi k/2}H^{(1)}_{ik}\left(\frac{p}{\omega}\,e^{-\omega t}\right) e^{i \vec{p}\cdot\vec{
x}}\,,\label{eq:sol}
\end{equation}
where $H^{1}_{\mu}(z)$ is a Hankel function of first kind, $p=|\vec{p}|$ is the momentum modulus and we note:
\begin{equation}
k=\sqrt{\mu^2-\textstyle{\frac{9}{4}}}\,, \quad \mu=\frac{m}{\omega}\,.
\label{eq:k}
\end{equation}
The fundamental solutions of negative
frequencies are $f_{\vec{p}}^*(x)$.
These solutions satisfy the ortonormalizations relations \cite{B12}:
\begin{eqnarray}
i\int d^3x\, (-g)^{1/2}\, f_{\vec{p}}^*(x)
\stackrel{\leftrightarrow}{\partial_{t}} f_{\vec{p}^{\,\,\prime}}(x)&=&\nonumber\\
-i\int d^3x\, (-g)^{1/2}\, f_{\vec{p}}\,(x)
\stackrel{\leftrightarrow}{\partial_{t}} f^*_{\vec{p}^{\,\,\prime}}(x)
&=&\delta^3(\vec{p}-\vec{p}^{\,\,\prime})\,,\\
i\int d^3x\, (-g)^{1/2}\, f_{\vec{p}}\,(x)
\stackrel{\leftrightarrow}{\partial_{t}} f_{\vec{p}^{\,\,\prime}}(x)&=&0\,,
\end{eqnarray}
where the integration extends on an arbitrary hypersurface
$t=const$ and $(-g)^{1/2}=e^{3\omega t}$,and
 the completeness condition \cite{B12}:
\begin{equation}
i\int d^3p\,  f^*_{\vec{p}}(t,\vec{x}\,) \stackrel{\leftrightarrow}{\partial_{t}}
f_{\vec{p}}\,(t,\vec{x}^{\,\,\prime})=e^{-3\omega t}\delta^3(\vec{x}-\vec{x}^{\,\,\prime})\,.\label{eq:comp}
\end{equation}

Thus we have now all the elements needed for beginning our study of Coulomb scattering for a scalar field on de Sitter space-time.

\section{\label{sec:3}Scattering amplitude}

In the well establish quantum field theory in Minkowski space, the scattering amplitudes can be calculated without ambiguity, and the main results are the conservation of energy and momentum in scattering processes. We want to see how these results will be affected by a curved background. This section will be entirely dedicated to the study of Coulomb scattering for a massive charged scalar field on de Sitter space.

It is known that the perturbation scheme used to solve interaction problems on Minkowski
spacetime can be successfully adapted to the curved spacetime \cite{B19}. It follows that the definition of the scattering amplitude in the first order of perturbation theory in scalar electrodynamics on de Sitter space will be:
\begin{equation}
A_{i\rightarrow f}=-e
\int\sqrt{-g(x)}\left[f^*_{\vec{p}^{\,\,\prime}}(x)\stackrel{\leftrightarrow}{\partial_{\mu}}f_{\vec{p} }\,(x)\right]A^{\mu}(x)d^{4}x,
\label{eq:ampl2}
\end{equation}
where the bilateral derivative acts as: $g\stackrel{\leftrightarrow}{\partial}h=(\partial h)g-(\partial g)h$.

In our calculation we need the form of the Coulomb potential on de
Sitter space-time which will depend on line element. Here the
situation is immediate if one recalls the conformal invariance of
the Maxwell equations and that the de Sitter metric is conformal
with the Minkowski one. If we write $\frac{Ze}{|\vec{x}|}$ for
the Coulomb field in the Minkowski space, then we find for the
corresponding de Sitter potential:
\begin{equation}
A_{\widehat{0}}(x)=A^{\widehat{0}}(x)=\frac{Ze}{|\vec{x}|}\,e^{-\omega t},\label{eq:pot}
\end{equation}
where the hatted indices indicate that we refer to the components
in the local Minkowski frames. We also observe that (\ref{eq:pot}) is just
the expression from flat space with distances dilated/contracted
by the factor $e^{-\omega t}$.

Having the above elements we can proceed to develop
the theory for the scattering amplitude on de Sitter spacetime. Here the
necessary requirements for developing the scattering theory is the
global hyperbolicity of the space \cite{B8} and having a complete set of
solutions of the free equation for the incident field and the scattered
field (Born approximation), with the distinction between positive
and negative frequencies.

Let $\phi_{i}(x)$ and $\phi_{f}(x)$ be the freely propagating
waves in the $in$ and $out$ sectors, and we assume that they are both
plane wave of positive frequency. Our intention is to
calculate the amplitude of Coulomb scattering
for the external field (\ref{eq:pot}) and for the initial and final states of
the form:
 \begin{equation}
\phi_{f}(x)=f_{\vec{p}_{f}}(x),\quad
\phi_{i}(x)=f_{\vec{p}_{i}}(x).
 \end{equation}
Replacing our quantities of interest (\ref{eq:pot}) and (\ref{eq:sol}) in the definition of scattering amplitude (\ref{eq:ampl2}) we observe that the dependence on $t, \vec{x}$ allows us
to split the four dimensional integral into a spatial
integral and a temporal one. The spatial integral have exactly the same form as in the Minkowski space \cite{B4}:
\begin{equation}
\int d^{3}x
\frac{e^{i(\vec{p_{i}}-\vec{p_{f}})\vec{x}}}{|\vec{x}|}=\frac{4
\pi}{|\vec{p_{f}}-\vec{p_{i}}|^{2}} .
\end{equation}
It is clear that the complicated part in our calculations is the temporal integral which contain the influence of the gravitational field via the factor $k$. Upon performing the double derivative and making use of the recurrence relation (\ref{eq:rh}) our amplitude that contains the temporal integrals can be brought to the form :
\begin{eqnarray}
 A_{i\rightarrow f}&=&
\frac{-\alpha Z}{8\pi|\vec{p_{f}}-\vec{p_{i}}|^{2}}
\biggl\{-\frac{p_{i}}{2}\int^{\infty}_{0}z\left[H^{(2)}_{ik}(p_{f}z)H^{(1)}_{ik-1}(p_{i}z)-H^{(2)}_{ik}(p_{f}z)H^{(1)}_{ik+1}(p_{i}z)\right]
dz\nonumber\\
&&+\frac{p_{f}}{2}\int^{\infty}_{0}z\left[H^{(2)}_{ik-1}(p_{f}z)H^{(1)}_{ik}(p_{i}z)-H^{(2)}_{ik+1}(p_{f}z)H^{(1)}_{ik}(p_{i}z)\right]
dz\biggl\},\label{eq:ampl3}
\end{eqnarray}
where $\alpha=e^{2}$ and we pass to a new variable of integration :
\begin{equation}
z=\frac{e^{-\omega t}}{\omega}.\label{eq:z}
\end{equation}
It is important to point out that in (\ref{eq:ampl3}) the integration limits for time variable corresponds to
$t=\pm\infty$ , assuming that the interaction extends into the
past and future.

The evaluation of the integrals is discussed in Appendix A, here we
give the final result in terms of: delta Dirac function $\delta$,
hypergeometric Gauss functions $_{2}F_{1}$ , beta Euler functions $B$ , gamma Euler functions $\Gamma$ and Heaviside step
function $\theta$:
\begin{eqnarray}
 A_{i\rightarrow f}&=&
\frac{- \alpha Z}{8\pi|\vec{p_{f}}-\vec{p_{i}}|^{2}}B_{k},\label{eq:ampl4}
\end{eqnarray}
where
\begin{eqnarray}
B_{k}&=&\frac{(p_{f}+p_{i})}{\sqrt{p_{f}p_{i}}}2i\delta(p_{f}-p_{i})+
\frac{1}{p_{i}}\theta(p_{i}-p_{f})\left[ h^*_{k}\left(\frac{p_{f}}{p_{i}}\right)
-g_{k}\left(\frac{p_{f}}{p_{i}}\right)\right]\nonumber\\
&&+\frac{1}{p_{f}}\theta(p_{f}-p_{i})\left[g^*_{k}\left(\frac{p_{i}}{p_{f}}\right)
-h_{k}\left(\frac{p_{i}}{p_{f}}\right)\right].\label{eq:b}
\end{eqnarray}
The above formulas are the main result. In Eq. (\ref{eq:b}) the notations are as follow:

\begin{eqnarray}
g_{k}(\chi)&=&\frac{e^{-\pi k}\chi^{ik}}{\sinh^{2}(\pi k)}\left[\frac{_{2}F_{1}\left(\frac{3}{2},\frac{1}{2}+ik;1+i
k;\chi^{2}\right)}{B(\frac{1}{2}-ik;1+ik)}-\frac{_{2}F_{1}\left(\frac{1}{2},\frac{3}{2}+ik;1+i
k;\chi^{2}\right)}{B(-\frac{1}{2}-ik;1+ik)}\right]\nonumber\\
&&+\frac{e^{\pi k}\chi^{-ik}}{\sinh^{2}(\pi k)}\left[\frac{_{2}F_{1}\left(\frac{3}{2},\frac{1}{2}-ik;1-i
k;\chi^{2}\right)}{B(\frac{1}{2}+ik;1-ik)}-\frac{_{2}F_{1}\left(\frac{1}{2},\frac{3}{2}-ik;1-i
k;\chi^{2}\right)}{B(-\frac{1}{2}+ik;1-ik)}\right]\nonumber\\
&&+\frac{\chi^{ik}}{\sinh^{2}(\pi k)}\left[\frac{_{2}F_{1}\left(\frac{3}{2},\frac{1}{2}+ik;1+i
k;\chi^{2}\right)}{B(-\frac{1}{2};1+ik)}-\frac{_{2}F_{1}\left(\frac{1}{2},\frac{3}{2}+ik;1+i
k;\chi^{2}\right)}{B(\frac{1}{2};1+ik)}\right]\nonumber\\
&&+\frac{\chi^{-ik}}{\sinh^{2}(\pi k)}\left[\frac{_{2}F_{1}\left(\frac{3}{2},\frac{1}{2}-ik;1-i
k;\chi^{2}\right)}{B(-\frac{1}{2};1-ik)}-\frac{_{2}F_{1}\left(\frac{1}{2},\frac{3}{2}-ik;1-i
k;\chi^{2}\right)}{B(\frac{1}{2};1-ik)}\right],\label{eq:gn}
\end{eqnarray}
\begin{eqnarray}
h_{k}(\chi)&=&\frac{e^{-\pi k}\chi^{-ik}}{\sinh^{2}(\pi k)}\left[\frac{_{2}F_{1}\left(\frac{3}{2},\frac{3}{2}-ik;2-i
k;\chi^{2}\right)}{B(-\frac{1}{2}+ik;2-ik)}\cdot\chi^{2}-\frac{_{2}F_{1}\left(\frac{1}{2},\frac{1}{2}-ik;-i
k;\chi^{2}\right)}{B(-ik;\frac{1}{2}+ik)}\right]\nonumber\\
&&+\frac{e^{\pi k}\chi^{ik}}{\sinh^{2}(\pi k)}\left[\frac{_{2}F_{1}\left(\frac{3}{2},\frac{3}{2}+ik;2+i
k;\chi^{2}\right)}{B(-\frac{1}{2}-ik;2+ik)}\cdot\chi^{2}-\frac{_{2}F_{1}\left(\frac{1}{2},\frac{1}{2}+ik;i
k;\chi^{2}\right)}{B(\frac{1}{2}-ik;ik)}\right]\nonumber\\
&&+\frac{\chi^{ik}}{\sinh^{2}(\pi k)}\left[\frac{_{2}F_{1}\left(\frac{1}{2},\frac{1}{2}+ik;i
k;\chi^{2}\right)}{B(\frac{1}{2};ik)}-\frac{_{2}F_{1}\left(\frac{3}{2},\frac{3}{2}+ik;2+i
k;\chi^{2}\right)}{B(-\frac{1}{2};2+ik)}\cdot\chi^{2}\right]\nonumber\\
&&+\frac{\chi^{-ik}}{\sinh^{2}(\pi k)}\left[\frac{_{2}F_{1}\left(\frac{1}{2},\frac{1}{2}-ik;-i
k;\chi^{2}\right)}{B(\frac{1}{2};-ik)}-\frac{_{2}F_{1}\left(\frac{3}{2},\frac{3}{2}-ik;2-i
k;\chi^{2}\right)}{B(-\frac{1}{2};2-ik)}\cdot\chi^{2}\right],\label{eq:hn}
\end{eqnarray}

where we note $\chi=\frac{p_{f}}{p_{i}}$ or $\frac{p_{i}}{p_{f}}$ .

The above formulas is our result and in what follows we will
discuss some of their physical consequences. Before that we must
state that the argument $\chi$ in relations (\ref{eq:gn}),(\ref{eq:hn}) is considered in
interval $0\leq\chi<1$ (see the argument in the $\theta$
functions). This is the domain of convergence of the
hypergeometric functions. This observation implies that in the terms that contain hypergeometric functions we must always consider that $p_{f}\neq p_{i}$ and from this we conclude that in contrast with the flat space case, here the law of conservation for modulus of momentum is lost. In the limit $p_{f}=p_{i}$ the integrals from our amplitude can be expressed in terms of hypergeometric functions which reduce in this limit to delta Dirac functions from (\ref{eq:b}), as we explain in the appendix.

Let us see what happens when we make the Minkowski limit of (\ref{eq:ampl4}).
The contributions involving functions $g_{k}(\chi),h_{k}(\chi),f_{k}(\chi)$ will vanish in the limit $k \rightarrow \infty$ which is just the part of the amplitude responsible for momentum non-conserving case. Only the delta terms will be effective in (\ref{eq:ampl4}), and we obtain:
\begin{equation}
\frac{- \alpha Z}{8\pi|\vec{p_{f}}-\vec{p_{i}}|^{2}}B_{\infty}=\frac{- \alpha Z(p_{f}+p_{i})}{8\pi|\vec{p_{f}}-\vec{p_{i}}|^{2}}\frac{2i\delta(p_{f}-p_{i})}{\sqrt{p_{f}p_{i}}}.
\label{eq:min}
\end{equation}
We know that the corresponding amplitude which can be obtained studying the scattering of a charged scalar field on constant field $Q/|\vec{x}|$ from the Minkowski case is:
\begin{equation}
\frac{- \alpha Z}{8\pi|\vec{p_{f}}-\vec{p_{i}}|^{2}}B_{M}=\frac{- \alpha Z(E_{f}+E_{i})}{8\pi|\vec{p_{f}}-\vec{p_{i}}|^{2}}\frac{4i\delta(E_{f}-E_{i})}{\sqrt{E_{f}E_{i}}},
\label{eq:delta}
\end{equation}
where $E_{f},E_{i}$ are the Minkowski energies and $E^{2}=p^{2}+m^{2}$.
To obtain the ratio between de Sitter/Minkowski amplitudes it is useful to consider the ultra relativistic case (the case of large momenta) $p\gg m, p\simeq E$ ,and using (\ref{eq:delta}), (\ref{eq:min}) we obtain:
\begin{equation}
\frac{\delta(p_{f}-p_{i})}{2\delta(E_{f}-E_{i})}=\frac{1}{2}.
\end{equation}
One expects that in the limit of a zero expansion of space $\omega \rightarrow 0$, to obtain the Minkowski amplitude, since in this limit the metric and Klein-Gordon equation take the form from Minkowski space. Our result may be a little surprising, but can be explained as follows. One finds after going back to the integration with respect to $t$ that for $t\rightarrow\infty$ the integrand vanishes as $e^{-\omega t}$,
while for $t\rightarrow-\infty$ becomes divergent. As one sees for $t\rightarrow-\infty$ the potential also becomes divergent. So the question is how come that we obtain a finite result, because a divergent behavior of the potential seems to be incompatible with a perturbational calculation from the beginning. The answer might be as follows: as one goes back in time, with the divergent behavior of the potential, the contraction of the space will cause a similar divergent behavior in the incident frequencies as one see from (\ref{eq:sol}), which leads finally to a finite result in our calculations.
Using now the asymptotic formulas for Hankel functions when $z$ takes large values \cite{B5}:
\begin{equation}
H^{(1,2)}_{\nu}(z)=\left(\frac{2}{\pi z}\right)^{1/2}e^{\pm
i(z-\nu\pi/2-\pi/4)}, z\rightarrow\infty,\label{eq:asym}
\end{equation}
and passing to conformal time $t_{c}=-z$ the
contributions to the amplitude (if we take the contributions of all integrals from (\ref{eq:ampl3})) will be:
\begin{equation}
\frac{(p_{f}+p_{i})}{\sqrt{p_{f}p_{i}}}\frac{1}{2\pi}\int dt_{c}e^{-i(p_{f}-p_{i})t_{c}}.
\label{eq:tc}
\end{equation}
 The integral in (\ref{eq:tc}) produces the distributional term, integrating
over an infinite interval. Observing that in (\ref{eq:tc}) appears only the
semi-infinite axis we obtain the 1/2 factor (a similar argument shows that the factor 1/2 is obtained from the last integral from (\ref{eq:ampl3}) too). To summarize, the
fact that we don't recover the Minkowski amplitude is hidden in
the expansion of space from infinite future because we don't have
contribution from infinite future in (\ref{eq:tc}). One should not
generally expect to recover Minkowski limit because the metric (\ref{eq:m})
is geodesically incomplete and  is conformal only to a part of
Minkowski space. Also is not surprising that the  momentum conservation is broken due
to the fact that the perturbation potential acts as an external time dependent source.

The result obtained in (\ref{eq:ampl4}) for scattering amplitude lead to an important physical consequence which is the fact that the modulus of the momentum is not conserved in the scattering process. Let us make some comments about this result.  Our conclusion is not so unexpected if one recalls that in Minkowski case the conservation of the momentum modulus is a direct consequence of the conservation of total energy which in turn is a consequence of the translational invariance with respect to time, which is lost in de Sitter space.

Let us make some comments about the cross section calculated with our amplitude.
In the theory of quantum fields from Minkowski space the problem of calculating cross section is solved defining the probabilities of transition in unit of time. The probability of transition in unit of time divided by incident flux is the cross section \cite{B7}. In our case the situation is complicated because our amplitude contains momentum non-conserving terms.  In Minkowski theory an essential role in the treatment when we deal with finitely extended wave is played by the energy conservation factor $\delta(E_{f}-E_{i})$. This factor assures that the cross section will not depend on the form of the incident wave, which clearly is not the case here. 

In addition another problem appears when one explicitly computes the incident flux.
The incident flux is obtained using the formula for the scalar current of particles:
\begin{equation}
J_{\mu}=-i\left[f^*_{\vec{p}}(x)\stackrel{\leftrightarrow}{\partial_{\mu}}f_{\vec{p}}\,(x)\right].\label{eq:flux}
\end{equation}
Using (\ref{eq:flux}) we obtain for the spatial component:
\begin{eqnarray}
j(t)=-i\left[f^*_{\vec{p}_{i}}(x)\stackrel{\leftrightarrow}{\partial_{i}}f_{\vec{p}_{i}}(x)\right]=\frac{\pi p_{i}e^{-3\omega t}}{2\omega (2\pi)^{3}}H^{(2)}_{ik}\left(\frac{p_{i}}{\omega}\,e^{-\omega t}\right)
H^{(1)}_{ik}\left(\frac{p_{i}}{\omega}\,e^{-\omega t}\right),\label{eq:flux1}
\end{eqnarray}
which is the expression of the incident flux. We see from (\ref{eq:flux1}) that this is a quantity that depends on time which is an unusual situation. Recalling now the fact that in the scattering theory the cross section is calculated with a flux independent of time, it is obviously that here we have to deal with a complicated problem. The fact that the incident flux is a quantity dependent of time together with the complicated form of the scattering probability leaves us with an ambiguity about how one can define the cross section in this case. Thus we are forced to accept that contrary to the well establish facts from flat space case, the cross section calculated with our amplitude and flux will be in general a quantity that depends on the form of the incident wave, which is a complex situation.

\section{\label{sec:5}Conclusion}

In this paper we investigated the Coulomb scattering amplitude for
the charged scalar field in an expanding de Sitter space. We have completely
ignored complications due to the ambiguity of the particle concept
in a curved space-time. We considered the initial and final states
of the field as exact solutions (with definite momentum) of the free Klein-Gordon equation in de
Sitter space. Our calculations shows that the amplitude depends in an essential way on the parameter
$k=\sqrt{\mu^2-\textstyle{\frac{9}{4}}}$. We also obtain that the modulus of momentum is not conserved in the scattering process on de Sitter space. In the limit of zero expansion of the space we don't recover the Minkowski amplitude, and we explain this as a consequence of the fact that our metric is geodesically incomplete. Another unexpected result is that the cross section is a quantity that depends on the form of the incident wave. All this unusual consequences are the result of the loss of translational invariance with respect to time on de Sitter space-time.

\appendix
\section{\label{sec:6}Integrals of the type Weber-Schafheitlin}

Here we give the main steps that help us to obtain the scattering amplitude (\ref{eq:ampl4}). The formulas
that we need for the first four integrals in (\ref{eq:ampl4}) are given in \cite{B5}, \cite{B6}:

\begin{eqnarray}
H^{(1)}_{\mu}(z)=\frac{J_{-\mu}(z)-e^{-i\pi\mu}J_{\mu}(z)}{i\sin(\pi\mu)}\nonumber\\
H^{(2)}_{\mu}(z)=\frac{e^{i\pi\mu}J_{\mu}(z)-J_{-\mu}(z)}{i\sin(\pi\mu)},
\end{eqnarray}
and the recurrence relation  \cite{B5}, \cite{B6}:
\begin{equation}
2\frac{dH^{(1,2)}_{\mu}(z)}{dz}=H^{(1,2)}_{\mu-1}(z)-H^{(1,2)}_{\mu+1}(z),
\label{eq:rh}
\end{equation}
which is used after we perform the bilateral derivation in our amplitude.
Replacing the above formulas in (\ref{eq:ampl3}) we arrive at
integrals of the Weber-Schafheitlin type \cite{B5}:
\begin{eqnarray}
\int^{\infty}_{0}dzz^{-s}J_{\mu}(a z)J_{\nu}(b z)&=&\frac{a^{\mu}\Gamma(\frac{\mu+\nu-s+1}{2})}{2^{s}b\,^{\mu-s+1}
\Gamma(\mu+1)\Gamma(\frac{\nu-\mu+s+1}{2})}\nonumber\\
&&\times_{2}F_{1}\left(\frac{\mu+\nu-s+1}{2},\frac{\mu-\nu-s+1}{2};\mu+1;\frac{a^{2}}{b^{2}}\right),\nonumber\\
&& b>a>0,Re(s)>-1,Re(\mu+\nu-s+1)>0,\label{eq:int1}
\end{eqnarray}
and
\begin{eqnarray}
\int^{\infty}_{0}dzz^{-s}J_{\mu}(a z)J_{\nu}(b z)&=&\frac{b^{\nu}\Gamma(\frac{\mu+\nu-s+1}{2})}{2^{s}a^{\nu-s+1}
\Gamma(\nu+1)\Gamma(\frac{\mu-\nu+s+1}{2})}\nonumber\\
&&\times_{2}F_{1}\left(\frac{\mu+\nu-s+1}{2},\frac{\nu-\mu-s+1}{2};\nu+1;\frac{b^{2}}{a^{2}}\right),
\nonumber\\
&&a>b>0,Re(s)>-1,Re(\mu+\nu-s+1)>0,\label{eq:int2}
\end{eqnarray}
we specify that the above two cases are fulfilled in our analysis. Our integrals demand  $s=-1$.
We will evaluate the integrals in (\ref{eq:ampl3}) directly
with $s=-1$ and with $\mu,\nu$ arguments following from (\ref{eq:k}).

Let us take a look to the distributional contribution
$\delta(p_{f}-p_{i})$. The distributional term
$\delta(a-b)$ is generally present in the integrals
(\ref{eq:int1}),(\ref{eq:int2}) for $s=-1$ (the case of real arguments $\mu=\nu$ is a
familiar one).

\textbf{Acknowledgements}

We would like to thank Professor Ion I.Cot\u aescu for reading the
manuscript and for useful suggestions that help us
to improve this work.

\end{document}